\documentclass[aip,rsi,groupedaddress,superscriptaddress,reprint]{revtex4-1}

\usepackage{graphicx}
\usepackage{dcolumn}
\usepackage{textcomp}
\usepackage{times}
\newcommand{\ket}[1]{|{#1}\rangle}

\begin{document}

\title{Preparation of an Exponentially Rising Optical Pulse for Efficient
  Excitation of Single Atoms in Free Space} 
\author{Hoang Lan Dao}
\affiliation{Centre for Quantum Technologies,
  National University of Singapore, 3 Science Drive 2,  Singapore, 117543}
\affiliation{Faculty of Science and Technology, University of Twente}
\author{Syed Abdullah Aljunid}
\affiliation{Centre for Quantum Technologies,
  National University of Singapore, 3 Science Drive 2,  Singapore, 117543}
\author{Gleb Maslennikov}
\affiliation{Centre for Quantum Technologies,
 National University of Singapore, 3 Science Drive 2,  Singapore, 117543}
\author{Christian Kurtsiefer}
\email[]{christian.kurtsiefer@gmail.com}
\affiliation{Centre for Quantum Technologies,
  National University of Singapore, 3 Science Drive 2,  Singapore, 117543}
\affiliation{Department of Physics,
  National University of Singapore, 2 Science Drive 3,  Singapore, 117542}

\date{\today}

\begin{abstract}
We report on a simple method to prepare optical pulses with exponentially rising
envelope on the time scale of a few ns. The scheme is based on the exponential
transfer function of a fast transistor, which generates an exponentially
rising envelope that is transferred first on a radio frequency carrier, and
then on a coherent cw laser beam with an electro-optical phase modulator
(EOM). The temporally shaped sideband is then extracted with an optical
resonator and can be used to efficiently excite a single $^{87}$Rb
atom.
\end{abstract}


\maketitle

\section{Introduction\label{intro}}
In recent years there has been significant progress towards establishing an
efficient quantum interface between photons and atoms. Such an interface is commonly
considered as a building block for complex quantum network circuits where atoms exchange
information (photons) via fundamental processes of absorption and emission~\cite{Monroe:09, Cirac:04, Cirac:97}.
For this processes to be efficient without any field enhancement by cavity
structures, there should be a significant overlap between atomic and 
photonic modes in spatial and time-frequency domains. In other words, the photons 
must be spatially and temporarily ``shaped'' to match the properties of atomic 
(electric dipole) transitions~\cite{Sondermann:07, Stobinska:09, Wang:11}.

For the spatial matching, the spatial profile of atomic absorption/emission
pattern is approximated by an oscillating (rotating) electric dipole,
corresponding to $\pi$ ($\sigma^+,\sigma^-$) transitions. An almost perfect
spatial overlap of light with a $\pi$ transition can be achieved with a
radially polarized light focused by parabolic mirror to an
atom~\cite{Sondermann:07}. However, a significant spatial overlap between a
Gaussian mode and $\sigma^+,\sigma^-$ transitions can already be accomplished
by focusing with commercially available high numerical aperture
lenses~\cite{Tey:09}.

For the temporal matching, a simple argument suggests that the time-reversal
of a spontaneous emission process as described by Weisskopf and Wigner,
namely an exponentially rising electrical field strength, should yield a high
excitation probability. It was shown in a few recent theoretical articles that
that for a given spatial overlap the highest excitation probability of an atom
occurs for photons with an exponentially rising envelope that is terminated by
a sharp drop~\cite{Stobinska:09, Wang:11}.

A straight-forward approach to generate coherent light pulses with an adequate
temporal shape uses some sort of optical modulator for a cw light beam, and
some electronic way to prepare an envelope signal that has an exponential
rise time on the order of a few nsec. While an exponentially decaying envelope
signal on this time scale is very simple to prepare e.g. by a RC low pass
filter, an exponential rising envelope is slightly more difficult. While these
days arbitrary function generators for analog signals have just about become
fast enough to implement this directly, we present a much simpler method in
this paper.

\section{Implementation\label{implementation}}

\begin{figure}
\begin{center}
\includegraphics[width=\columnwidth]{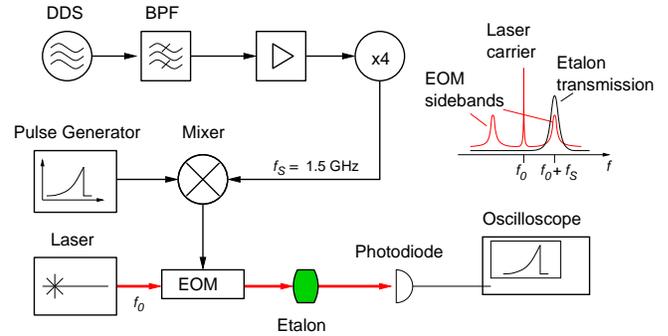}
\caption{\label{expsetup}
Experimental setup. A signal at 375\,MHz is extracted with a band pass
filter (BPF) from a direct digital synthesizer (DDS) and frequency-quadrupled
to $f_S=$1.5\,GHz. This RF carrier is modulated with the exponentially shaped
envelope function on a mixer, and sent to an electro-optical modulator (EOM)
that generates optical sidebands around an optical center frequency $f_0$. An
etalon extracts one of the RF sidebands. The optical pulse is monitored with a
a fast photodiode.}
\end{center}
\end{figure}

An overview of our experimental setup is shown in FIG~\ref{expsetup}. The core
idea is to use the transfer function of a transistor to convert a linear 
rising electrical signal into an exponentially rising current. 
With this current, we first modulate the envelope of a radio frequency carrier,
which is then used in an electro-optical phase modulator (EOM) to imprint
optical
side bands onto a coherent, frequency-stabilized cw laser beam. An etalon is
used to filter out one of the sidebands with enough bandwidth to recover the
exponential shape in time, and sent to a fast photodiode for characterization.
In the following, we describe the various building blocks in more detail.

\subsection{Electrical pulse\label{elec}}
\begin{figure}
\begin{center}
\includegraphics[width=0.9\columnwidth]{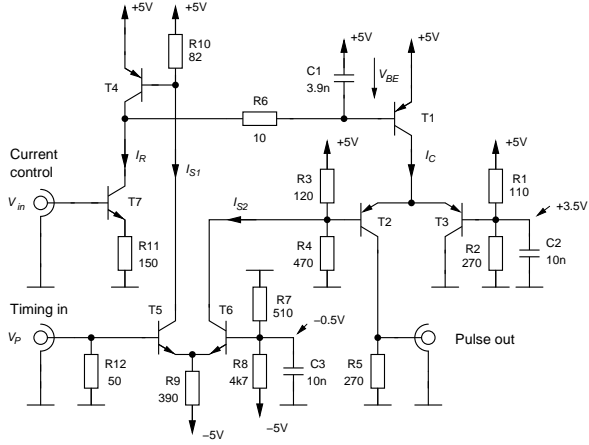}
\caption{\label{pulse_circuit}
Electrical circuit for exponential pulse generation. T1-T4 are wide bandwidth PNP transistors
(BFT93),  T5-T7  wideband NPN transistors (BFR93). See text for details of 
operation.}
\end{center}
\end{figure}

The collector current $I_C$ of an unsaturated bipolar junction transistor
depends on its base voltage $V_{BE}$ according Shockley relationship: 

\begin{equation}
I_C=I_0\times (\exp(\frac{V_{BE}}{V_T})-1)\,,
\end{equation}

where $I_0$ is the reverse saturation current and
$V_T=kT/e\approx26$\,mV at room temperature. 
If $V_{BE}$ increases linearly in time, the collector current 
grows in good approximation exponentially if
$I_C \gg I_0$. Figure~\ref{pulse_circuit} shows the electrical circuit that
implements this idea, together with other components ensuring a fast
switch-off. The design time constant $\tau_R$ for the exponential rise was about 27\,ns
to match the decay time of the optical transition on a D2 line in Rubidium.

A linearly rising $V_{BE}$ is provided by charging capacitor C1  with
a constant current $I_R$.
Transistor T1 then performs the transformation of the linear slope into an
exponentially rising  current $I_C$. For the nominal
$\tau_R=27$\,ns a slope  $\partial V_{BE}/\partial t\approx10^6$\,V/s is
necessary. For C1=3.9\,nF, this slope can be accomplished with a reasonable
charging current $I_R=3.9$\,mA. 

The charge current $I_R$ is provided by the current source combination T7 and
R11, which 
generates a current defined by an analog input voltage $V_\mathrm{in}$, and allows for a
variation of $\tau_R$ by a factor of about 5 in both directions for exploring
different interaction regimes of the optical pulse with the atom. The
exponential time constant of the output pulse is then roughly given by
\begin{equation}
  \tau_R=\mathrm{R11}\, \mathrm{C1}\,{\,V_T\over(V_\mathrm{in}-0.7\,V)}\,.
\end{equation}

The desired pulse has not only an exponential rise, but also a steep cutoff at
a given time, and the whole shape of the pulse needs to be defined with
respect to some external timing reference. For this purpose, we use a digital
signal following a NIM standard suitable to interact with our control
equipment. This timing signal starts the charging of C1 when active
($V_P=-1$\,V), and routes the the output current $I_C$  via T2 into the load
impedance. When switching to passive state ($V_P\approx 0$\,V), the output
current $I_C$ is diverted through T3 away from the output, 
and C1 is discharged via T4.  The basis voltage levels of T2,T3 are chosen such
that the main transistor T1 has a collector potential of 3.7...4.2\,V to keep
it unsaturated. T2 and T3 themselves stay out of
saturation for an output voltage up to 2V corresponding to $I_C\approx40$\,mA.

\begin{figure}
\begin{center}
\includegraphics[width=0.9\columnwidth]{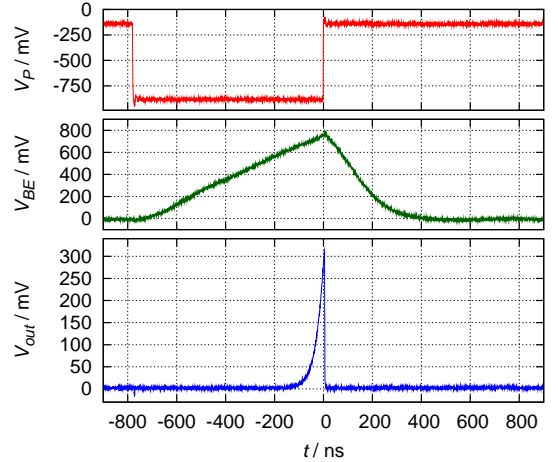}
\caption{\label{pulse_traces}
Operation of the pulse shaper. Triggered by $V_P$ (top trace), the
base-emitter voltage of transistor T1 rises linearly (middle trace), leading to
an exponential rising voltage $V_\mathrm{out}$ (bottom trace). When the
trigger signal 
returns to passive, the current is diverted away from the output, leading to
a sudden drop of the envelope, while capacitor C1 is slowly discharged.}
\end{center}
\end{figure}

The timing of the pulse is now critically determined by the length of the
control pulse $T_P$, and critically dependent on the temperature voltage as
well. A timing diagram of the relevant voltages for a typical time constant of
30.5\,ns is shown in figure~\ref{pulse_traces}.

\subsection{RF generation and mixing\label{rf}}

In the experiment, the resulting optical frequency of the pulse has to be tuned across the 
closed cycling transition of $^{87}\mathrm{Rb}$ which occurs between hyperfine states of D2 line
$\ket{\mathrm{F}=2}\rightarrow \ket{\mathrm{F'}=3}$ at $f_\mathrm{opt}\approx384$\,THz.
The RF carrier at frequency $f_S$ that is modulated with the exponential
envelope signal defines the splitting between the optical carrier frequency of the
laser and the sidebands that are obtained with EOM.  We choose
$f_S=1.5$\,GHz, because in this case, $f_0+f_S=f_\mathrm{opt}$ hits the transition
frequency of D2 line in $^{85}\mathrm{Rb}$ between $\ket{\mathrm{F}=3}$ and
$\ket{\mathrm{F'}=4}$, and the laser carrier frequency $f_0$ can be locked to
a vapor cell containing the natural isotope distribution of Rb.

The RF carrier is prepared by a direct digital synthesizer (DDS)
based on an Analog Devices AD9958 chip, running at a sampling frequency of
500\,MHz. From this, we reconstruct the first upper mirror frequency at
375\,MHz with a strip line band pass filter (spurious 
suppression: 70\,dBc at 125\,MHz, 24\,dBc at 500\,MHz, 35\,dBc at
625\,MHz). After an amplification to about 0\,dbm this carrier is twice frequency-doubled
(Mini-circuits AMK2-13 and KC2-11) to the desired $f_S=1.5$\,GHz, and used as a local
oscillator input of a double-balanced mixer (Mini-circuits ADE-30). The
attenuated envelope signal is connected to the IF port of the
mixer.

\begin{figure}
\begin{center}
\includegraphics[width=0.9\columnwidth]{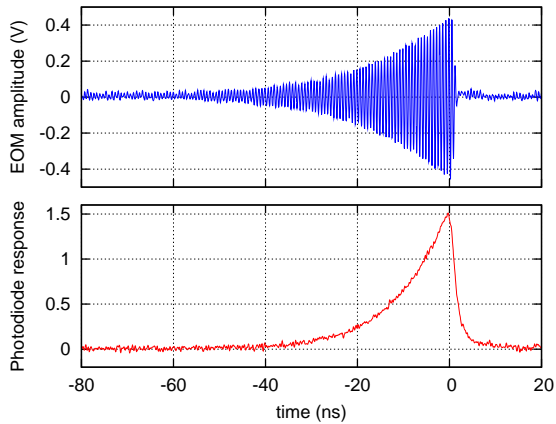}
\caption{\label{fig:RFandPIN}
Modulated carrier at $f_S=1.5$\,GHz with the exponential envelope (upper
trace), and optical response after the etalon (lower trace). The different
exponential decay time constants result from the squaring of the optical
amplitude under power detection by the pin diode.
}
\end{center}
\end{figure}

\subsection{EOM and filter cavities \label{eom}}
The modulated RF signal around $f_S$ is converted into an optical sideband
with a waveguide electro-optical phase modulator (EOSpace, model
PM-0K5-10-PFA-PFA-780). It has an electrical bandwidth of nominally 20\,GHz
and a half wave voltage $V_\pi=1700$\,mV. Since the amplitude of the optical
side band is proportional to $J_1(\pi V_\mathrm{RF}/V_\pi)$, where $J_1$ is
the Bessel function and $V_\mathrm{RF}$ the amplitude of the radio frequency
signal, $V_\mathrm{RF}$ must be kept low enough to minimize distortion of the
amplitude transfer due to the nonlinearity of the Bessel function. For
$V_\mathrm{RF}=0.1V_\pi$, the distortion is about 1.2\%, for $V_\mathrm{RF}=0.2V_\pi$ already 4.8\%. 

In a next step, one of the optical side bands needs to be isolated from the
optical carrier and the other sideband. We do this with an etalon that is
formed by a plano-convex fused silica substrate with a radius of curvature of
$R=50$\,mm and a center thickness of 6.35\,mm. Both surfaces of this lens
substrate are coated with dielectric mirrors with a reflectivity
of 95\%, leading to a free spectral range of about 17\,GHz and a transmission
line width (FWHM) of 273\,MHz. This solid etalon can be temperature-tuned
to transmit one optical sideband, with a temperature tuning of 7.4\,K for
one full spectral range of 17\,GHz. We keep the temperature stable to about
5\,mK, resulting in a frequency uncertainty of 11.5\,MHz. For the given etalon
reflectivity, there is still a transmission of 0.77\% at the optical carrier
frequency. We thus use three consecutive etalons to achieve more than 60 dB
carrier power extinction. 

A quick characterization of the emerging pulse was carried out with a fast PIN
silicon photodiode (Hamamatsu S5973, nominal bandwidth 1\,GHz) connected to a
oscilloscope with 2\,GHz bandwidth (see FIG~\ref{fig:RFandPIN}). While the RF
signal sent to the EOM shows an exponentially rising envelope with a time
constant of about 17.4\,ns, the optical pulse seen by the photodiode
shows a rise with a time constant of about 10\,ns. The time
constants should differ by a factor of 2 exactly due to the square dependency
of the optical power sensed by the photodiode from the electrical field
amplitude. Compared to the sudden drop in the modulated RF signal, the optical
response also shows a slower decay. This is partly due to the finite bandwidth
of the filter cavities, which induces a ringdown time of the triple cavity
system, and partly due to the photodiode response.  To realize a faster stop
of the optical wave packet, one would need to choose a larger filter cavity
bandwidth, and consequently a larger RF carrier frequency.

\section{Summary} 

We presented a very simple scheme to generate a fourier-limited
optical pulse with an exponentially rising amplitude and a sharp decay from a
cw laser tuned to a resonance of an atomic transition. For a large overlap of
the spatial mode of such a light field with the emission pattern of a dipole
transition, such a pulse should lead to high excitation probability of an atom
(or any other two-level system) with a small average photon number. 


\begin{thebibliography}{7}
\expandafter\ifx\csname natexlab\endcsname\relax\def\natexlab#1{#1}\fi
\expandafter\ifx\csname bibnamefont\endcsname\relax
  \def\bibnamefont#1{#1}\fi
\expandafter\ifx\csname bibfnamefont\endcsname\relax
  \def\bibfnamefont#1{#1}\fi
\expandafter\ifx\csname citenamefont\endcsname\relax
  \def\citenamefont#1{#1}\fi
\expandafter\ifx\csname url\endcsname\relax
  \def\url#1{\texttt{#1}}\fi
\expandafter\ifx\csname urlprefix\endcsname\relax\def\urlprefix{URL }\fi
\providecommand{\bibinfo}[2]{#2}
\providecommand{\eprint}[2][]{\url{#2}}

\bibitem[{\citenamefont{Luo et~al.}(2009)\citenamefont{Luo, Hayes, Hanning,
  Matsukevich, Maunz, Olmschenk, Sterk, and Monroe}}]{Monroe:09}
\bibinfo{author}{\bibfnamefont{I.~L.} \bibnamefont{Luo}},
  \bibinfo{author}{\bibfnamefont{D.}~\bibnamefont{Hayes}},
  \bibinfo{author}{\bibfnamefont{T.~A.} \bibnamefont{Hanning}},
  \bibinfo{author}{\bibfnamefont{D.~N.} \bibnamefont{Matsukevich}},
  \bibinfo{author}{\bibfnamefont{P.}~\bibnamefont{Maunz}},
  \bibinfo{author}{\bibfnamefont{S.}~\bibnamefont{Olmschenk}},
  \bibinfo{author}{\bibfnamefont{J.~D.} \bibnamefont{Sterk}}, \bibnamefont{and}
  \bibinfo{author}{\bibfnamefont{C.}~\bibnamefont{Monroe}},
  \bibinfo{journal}{Fortschritte der Physik} \textbf{\bibinfo{volume}{57}},
  \bibinfo{pages}{1133} (\bibinfo{year}{2009}).

\bibitem[{\citenamefont{Cirac and Zoller}(2004)}]{Cirac:04}
\bibinfo{author}{\bibfnamefont{J.~I.} \bibnamefont{Cirac}} \bibnamefont{and}
  \bibinfo{author}{\bibfnamefont{P.}~\bibnamefont{Zoller}},
  \bibinfo{journal}{Physics Today} \textbf{\bibinfo{volume}{57}},
  \bibinfo{pages}{38} (\bibinfo{year}{2004}).

\bibitem[{\citenamefont{Cirac et~al.}(1997)\citenamefont{Cirac, Zoller, Kimble,
  and Mabuchi}}]{Cirac:97}
\bibinfo{author}{\bibfnamefont{J.~I.} \bibnamefont{Cirac}},
  \bibinfo{author}{\bibfnamefont{P.}~\bibnamefont{Zoller}},
  \bibinfo{author}{\bibfnamefont{H.~J.} \bibnamefont{Kimble}},
  \bibnamefont{and} \bibinfo{author}{\bibfnamefont{H.}~\bibnamefont{Mabuchi}},
  \bibinfo{journal}{Phys. Rev. Lett.} \textbf{\bibinfo{volume}{78}},
  \bibinfo{pages}{3221} (\bibinfo{year}{1997}).

\bibitem[{\citenamefont{Sondermann et~al.}(2007)\citenamefont{Sondermann,
  Maiwald, Konermann, Lindlein, Peschel, and Leuchs}}]{Sondermann:07}
\bibinfo{author}{\bibfnamefont{M.}~\bibnamefont{Sondermann}},
  \bibinfo{author}{\bibfnamefont{R.}~\bibnamefont{Maiwald}},
  \bibinfo{author}{\bibfnamefont{H.}~\bibnamefont{Konermann}},
  \bibinfo{author}{\bibfnamefont{N.}~\bibnamefont{Lindlein}},
  \bibinfo{author}{\bibfnamefont{U.}~\bibnamefont{Peschel}}, \bibnamefont{and}
  \bibinfo{author}{\bibfnamefont{G.}~\bibnamefont{Leuchs}},
  \bibinfo{journal}{Appl. Phys. B} \textbf{\bibinfo{volume}{89}},
  \bibinfo{pages}{489} (\bibinfo{year}{2007}).

\bibitem[{\citenamefont{Stobi\'nska et~al.}(2009)\citenamefont{Stobi\'nska,
  Alber, and Leuchs}}]{Stobinska:09}
\bibinfo{author}{\bibfnamefont{M.}~\bibnamefont{Stobi\'nska}},
  \bibinfo{author}{\bibfnamefont{G.}~\bibnamefont{Alber}}, \bibnamefont{and}
  \bibinfo{author}{\bibfnamefont{G.}~\bibnamefont{Leuchs}},
  \bibinfo{journal}{Europhys. Lett.} \textbf{\bibinfo{volume}{86}},
  \bibinfo{pages}{14007} (\bibinfo{year}{2009}).
  \urlprefix\url{http://arxiv.org/abs/0808.1666v2}.

\bibitem[{\citenamefont{Wang et~al.}(2011)\citenamefont{Wang, Minar, Sheridan,
  and Scarani}}]{Wang:11}
\bibinfo{author}{\bibfnamefont{Y.}~\bibnamefont{Wang}},
  \bibinfo{author}{\bibfnamefont{J.}~\bibnamefont{Minar}},
  \bibinfo{author}{\bibfnamefont{L.}~\bibnamefont{Sheridan}}, \bibnamefont{and}
  \bibinfo{author}{\bibfnamefont{V.}~\bibnamefont{Scarani}},
  \bibinfo{journal}{Phys. Rev. A} \textbf{\bibinfo{volume}{83}},
  \bibinfo{pages}{063842} (\bibinfo{year}{2011}).

\bibitem[{\citenamefont{Tey et~al.}(2009)\citenamefont{Tey, Aljunid, Huber,
  Chng, Chen, Maslennikov, and Kurtsiefer}}]{Tey:09}
\bibinfo{author}{\bibfnamefont{M.~K.} \bibnamefont{Tey}},
  \bibinfo{author}{\bibfnamefont{S.~A.} \bibnamefont{Aljunid}},
  \bibinfo{author}{\bibfnamefont{F.}~\bibnamefont{Huber}},
  \bibinfo{author}{\bibfnamefont{B.}~\bibnamefont{Chng}},
  \bibinfo{author}{\bibfnamefont{Z.}~\bibnamefont{Chen}},
  \bibinfo{author}{\bibfnamefont{G.}~\bibnamefont{Maslennikov}},
  \bibnamefont{and}
  \bibinfo{author}{\bibfnamefont{C.}~\bibnamefont{Kurtsiefer}},
  \bibinfo{journal}{New Journal of Physics} \textbf{\bibinfo{volume}{11}},
  \bibinfo{pages}{043011} (\bibinfo{year}{2009}).

\end{thebibliography}

\end{document}